\documentclass{INTERSPEECH2023}

\interspeechcameraready

\usepackage{amsmath,graphicx}
\usepackage{adjustbox}
\usepackage[table,xcdraw]{xcolor}

\usepackage{lineno,hyperref}
\usepackage{enumerate}
\usepackage{setspace,url,float}
\usepackage{lscape,multicol,multirow,longtable,tabularx,array}
\newcolumntype{?}{!{\vrule width 0.8pt}}
\usepackage{tablefootnote}
\usepackage{subcaption}
\usepackage{makecell}

\usepackage{tabularx,arydshln}

\usepackage{array,etoolbox}
\preto\tabular{\setcounter{magicrownumbers}{0}}
\newcounter{magicrownumbers}

\title{Model ADaptation for ASR in low-resource Indian languages
(MADASR)}
\name{Abhayjeet Singh$^1$, Arjun Singh Mehta, Ashish Khuraishi K S, Deekshitha G, Gauri Date, Jai Nanavati, Jesuraja Bandekar, Karnalius Basumatary, Karthika P, Sandhya Badiger, Sathvik Udupa, Saurabh Kumar, Savitha, Prasanta Kumar Ghosh, Prashanthi V, Priyanka Pai, Raoul Nanavati, Rohan Saxena, Sai Praneeth Reddy Mora, Srinivasa Raghavan}
\address{
  $^1$Department of Electrical Engineering, Indian Institute of Science (IISc), Bangalore-560012, India\\
  $^2$Navana Tech (NT), Nanavati Mahalaya 18, Homi Mody St. Fort, Mumbai-400001, India}
\email{challenge.respin@iisc.ac.in}
\begin{document}
\maketitle
 
\begin{abstract}
Automatic speech recognition (ASR) performance has improved drastically in recent years, mainly enabled by self-supervised learning (SSL) based acoustic models such as wav2vec2 and large-scale multi-lingual training like Whisper. A huge challenge still exists for low-resource languages where the availability of both audio and text is limited. This is further complicated by the presence of multiple dialects like in Indian languages. However, many Indian languages can be grouped into the same families and share the same script and grammatical structure. This is where a lot of adaptation and fine-tuning techniques can be applied to overcome the low-resource nature of the data by utilising well-resourced similar languages.

In such scenarios, it is important to understand the extent to which each modality, like acoustics and text, is important in building a reliable ASR. It could be the case that an abundance of acoustic data in a language reduces the need for large text-only corpora. Or, due to the availability of various pretrained acoustic models, the vice-versa could also be true. In this proposed special session, we encourage the community to explore these ideas with the data in two low-resource Indian languages of Bengali and Bhojpuri. These approaches are not limited to Indian languages, the solutions are potentially applicable to various languages spoken around the world.

\end{abstract}

\section{Introduction}
Voice-based solutions (speech-to-text, text-to-speech, and natural language processing) are a necessity for digital technologies to be adopted in India because a large majority of users can communicate only through their spoken languages and dialects due the literacy constraints. With such a consideration, voice-based systems in Indian languages can potentially help illiterate Indians to use digital services by simply communicating in their own dialects. Regional variations in a language having some unique phonology and pronunciation characteristics are considered dialects in that language \cite{arabic_nmt_2018, phd_thesis_german_swiss}. Dialectal variation also includes variations in vocabulary and grammar. One dialect is confined to a particular region and can even be treated as a separate language \cite{phd_thesis_german_swiss}. 
Even though it is challenging to access and collect such dialect-specific resources, dialect-based analysis and translation systems help to mitigate the variabilities within a language. 

Bengali is the official language of West Bengal, a state in the eastern region of India \cite{benga}. It is the fourth most populous state that primarily depends on agriculture and medium-sized industry. Bengali (Bangla) is the second most widely spoken language in India, after Hindi with approximately 97 million native speakers (as per Census of India report, 2011) \cite{ldc}. It is the fifth most-spoken native language and the sixth most-spoken language in terms of the number of speakers in the world \cite{eth}. Bengali is spoken over the whole of West Bengal, Tripura, and some parts of Bihar, Orissa, and Assam. While the people of Bengal primarily speak Bengali, it is also spoken by Bengali immigrants in several countries around the world, with the greatest population in Bangladesh, Pakistan, Saudi Arabia, and UAE \cite{LS}. 

Whereas Bhojpuri is not considered an official language by the Indian constitution, as per the 2011 Linguistic Survey of India. But there are 505 lahks (50M) Bhojpuri speakers in India, mainly across the states of Uttar Pradesh and Bihar \cite{census2011}. According to survey reports, Bhojpuri is an Indo-Aryan language that is considered a dialect of Hindi~\cite{ldc, 6085979}. Even though both languages use the same written script - Devanagari, Bhojpuri is quite different from Hindi. A lot of work is done with the large resources available for Hindi, but there is a lack of focused studies on its variants. Because of this, Bhojpuri is rarely explored in speech research. Hence, based on the literature and resources available, Bhojpuri remains a low-resource language to date.

Speech and language technologies for Indian languages have been less explored compared to that for English. Different research works in the Bengali language include corpora collection \cite{ldc, ramesh2021samanantar, 6121518}, speech recognition \cite{6085979, csr, mridha2021challenges}, document classification \cite{HOSSAIN2021115394, karim2020classification}, sentiment analysis \cite{karim2020classification}, hate speech detection \cite{karim2020classification}, rule-based dialect translation \cite{10.1007/978-981-10-6890-4_44}, and  Bengali-English machine translation \cite{hasan2020lowresource, 10.5555/1621431.1621444}. Similarly, there has been an effort to study ASR in Bhojpuri \cite{kumar2022annotated}, though the dataset is of a small scale. And there exists a BHLTR Bhojpuri text corpus~\cite{ojha2019english} which has text scraped from Bhojpuri books, magazines, and websites.

Corpora is the backbone of the development of speech and language technologies \cite{6085979}, where the text data needed to be balanced in terms of dialects, whereas speech recordings needed to be balanced in terms of age, accent, and gender. Samanantar is the largest publicly available parallel corpora collection across Indic languages \cite{ramesh2021samanantar}. However, no such collection has been done across dialects within an Indian language be it either for automatic speech recognition (ASR) or natural language processing (NLP) applications. Putting an effort to collect dialect-specific text data will help in creating a dialect-specific speech corpus and language model which finally can be used for the development of an ASR.

Based on the geographical area, and the boundary of the neighboring states and countries, Bengali language can be classified into more than 6 dialects \cite{goo_book}. But in this work,  we focus only on five of these dialects, selected to cover the majority of the Bengali-speaking population. Table \ref{tab_dia} summarises the details of the five dialects under consideration. As per the 2011 census of India report, among the 97 million Bengali-speaking population, 96 million are following the standard variety of Bengali-SCB  whereas only 4 lakh are speaking RAJ \cite{census2011}. VAR of North Bengal, JAR of South and Lower  Bengal, and RAJ are the other dialects considered in the work. On a dialect-based survey done by us with 128 Bengali speakers, the number of persons having a dialect knowledge was in the order of 88, 14, 8, 13, and 5, respectively, for KOL, PUR, RAJ, JAR, and VAR dialects. 
\vspace{-0.15cm}
\begin{table}[h]
    \centering
    \footnotesize
    \begin{tabular}{|l|l|l|l|}
    \hline
    \textbf{Symbol}&\textbf{Code}& \textbf{Dialect}& \textbf{Districts}\\
    \hline
        D1&PUR&Western Bengali&
    Purba \& Paschim Medinipur\\
    D2&VAR&Varendri/Pundra/ &
    Malda, Dakshin Dinajpur\\
    &&Northern Bengali&\\
    D3&KOL&SCB&Kolkata\\
    D4&JAR&Jharkhandi&Purulia \& Bankura\\
    D5&RAJ&Rajbangshi &Jalpaiguri\\
    \hline
    \multicolumn{3}{l}{\footnotesize{SCB: Standard Colloquial Bengali}}
    \end{tabular}
    \vspace{-0.2cm}
    \caption{Details of the five Bengali dialects used in this work}
    \vspace{-0.2cm}
    \label{tab_dia}
\end{table}

\begin{table}[h]
\footnotesize
 \centering
 \begin{tabular}{|c|c|c|c|}
 \hline
 \textbf{Symbol}    &\textbf{Dialect} &\textbf{District}  &\textbf{State}\\
 \hline
 \multirow{2}{*}{D1}  &\multirow{2}{*}{NBH}   &Deoria&UP\\
 &&East Champaran &Bihar\\ \hline 
 D2  &WBH&Varanasi       &UP\\\hline
 D3  &SBH&Saran          &Bihar\\
 \hline
 \end{tabular}
 \caption{Details of regions focused to collect the dialect-rich Bhojpuri corpus}
 \label{bhoj}
 \vspace{-0.5cm}
\end{table}

In this work, we utilise in-house corpora of $1100$ hours of Bhojpuri data collected by targeting specific dialects. Based on the geographical area, the Bhojpuri language can be classified into $3$ major dialects as summarized in Table~\ref{bhoj}. These three major dialects (NBH-Northern Bhojpuri, WBH-Western Bhojpuri and SBH-Southern Bhojpuri) cover the majority of the Bhojpuri-speaking population in India.

\section{Datasets}

\section{Conclusion and Future work}
This helps to improve the reach, usage, and usefulness of voice-based digital technologies among the illiterate community. By identifying and analysing sentences that reflect the nuances of Bengali dialects, the technology ecosystem will be able to build models with the voice as a primary form of digital interaction and encourage the use of digital services to all in the near future. 

\section{Acknowledgements}
We thank everyone who supported the throughout this study. We especially thank the validators and other volunteers who contributed to the curation of Bengali and Bhojpuri datasets.

\bibliographystyle{IEEEtran}
\bibliography{mybib}

\end{document}